\documentclass[fleqn,usenatbib]{mnras}

\usepackage{newtxtext,newtxmath}

\usepackage[T1]{fontenc}
\usepackage{ae,aecompl}

\usepackage{graphicx}	
\usepackage{amsmath}	
\usepackage{amssymb}	

\newcommand{\Ni}{\ensuremath{^{56}\mathrm{Ni}}}
\newcommand{\Co}{\ensuremath{^{56}\mathrm{Co}}}
\newcommand{\Fe}{\ensuremath{^{56}\mathrm{Fe}}}

\newcommand{\Ein}{\ensuremath{E_\mathrm{in}}}
\newcommand{\Msun}{\ensuremath{\mathrm{M}_\odot}}
\newcommand{\Rsun}{\ensuremath{\mathrm{R}_\odot}}

\newcommand{\Eacc}{\ensuremath{\dot{E}_\mathrm{acc}}}

\title[Fallback accretion powered OGLE14-073]{OGLE-2014-SN-073 as a fallback accretion powered supernova}

\author[T. J. Moriya et al.]{
Takashi J. Moriya,$^{1}$\thanks{E-mail: takashi.moriya@nao.ac.jp (TJM)}\thanks{NAOJ Fellow}
Giacomo Terreran,$^{2,3}$
and Sergei I. Blinnikov$^{4,5,6}$
\\
$^{1}$Division of Theoretical Astronomy, National Astronomical Observatory of Japan, 
National Institutes of Natural Sciences, \\ 2-21-1 Osawa, Mitaka, Tokyo 181-8588, Japan\\
$^{2}$Center for Interdisciplinary Exploration and Research in Astrophysics (CIERA) and Department of Physics and Astronomy,\\ Northwestern University, Evanston, IL 60208\\
$^{3}$INAF-Osservatorio Astronomico di Padova, Vicolo dell'Osservatorio 5, 35122 Padova, Italy\\
$^{4}$Institute for Theoretical and Experimental Physics, Bolshaya Cheremushkinskaya ulitsa 25, 117218 Moscow, Russia \\
$^{5}$All-Russia Research Institute of Automatics, Sushchevskaya ulitsa 22, 127055 Moscow, Russia \\
$^{6}$Kavli Institute for the Physics and Mathematics of the Universe (WPI),
The University of Tokyo Institutes for Advanced Study, \\
The University of Tokyo, 5-1-5 Kashiwanoha, Kashiwa, Chiba 277-8583, Japan \\
}

\date{Accepted 2017 December 7. Received 2017 December 7; in original form 2017 October 27}

\pubyear{2017}

\begin{document}
\label{firstpage}
\pagerange{\pageref{firstpage}--\pageref{lastpage}}
\maketitle

\begin{abstract}
We investigate the possibility that the energetic Type~II supernova OGLE-2014-SN-073 is powered by a fallback accretion following the failed explosion of a massive star. Taking massive hydrogen-rich supernova progenitor models, we estimate the fallback accretion rate and calculate the light curve evolution of supernovae powered by the fallback accretion. We find that such fallback accretion powered models can reproduce the overall observational properties of OGLE-2014-SN-073. It may imply that some failed supernovae could be observed as energetic supernovae like OGLE-2014-SN-073 instead of faint supernovae as previously proposed.
\end{abstract}




\section{Introduction}
Core-collapse supernovae (SNe) are explosions of massive stars that exceed around 10~\Msun\ at the zero-age main sequence (ZAMS). Many of them retain hydrogen at the time of the explosions and they are observed as Type~II SNe. Most Type~II SNe are found to have an explosion energy of around $10^{51}~\mathrm{erg}$ \citep[e.g.,][]{pejcha2015snii} that can be explained by the standard neutrino-driven explosion mechanism \citep[e.g.,][]{janka2012corecollapsereview}.

However, recent transient surveys are starting to find energetic Type~II SNe that do not fit to the canonical picture. For example, some superluminous SNe are found to have spectral signatures of hydrogen and their explosion energies might be well above $10^{51}~\mathrm{erg}$ \citep[e.g.,][]{moriya2013sn2006gy}. The particularly bright Type~II SN~2009fk \citep{botticella2010sn2009kf} is also suggested to have large explosion energies \citep{utrobin2010sn2009kf}, although its high luminosity may alternatively be related to the interaction of the ejecta with a dense circumstellar medium \citep{moriya2011rsgwindexp}.
 
\citet{terreran2017ogle73} recently presented the Type~II SN OGLE-2014-SN-073 (OGLE14-073 hereafter) that clearly exceeds the canonical explosion energy of $10^{51}~\mathrm{erg}$. Based on the well-sampled light curve (LC) and spectra, they conclude that OGLE14-073 has the explosion energy of $12.4^{+13.0}_{-5.9}\times 10^{51}~\mathrm{erg}$ with the ejecta mass of $60^{+42}_{-16}~\Msun$. Such a huge explosion energy is well beyond what can be provided by the neutrino-driven mechanism and alternative sources of energy must be invoked in order to explain the observables of OGLE14-073. \citet{terreran2017ogle73} raise several possible explosion scenarios to explain OGLE14-073 including pair-instability SNe \citep[e.g.,][]{barkat1967pisn,rakavy1967pisn}, pulsational pair-instability SNe \citep[e.g.,][]{woosley2017ppisn}, and magnetar-powered hydrogen-rich SNe \citep[e.g.,][]{bersten2016hrichmag,sukbold2017hrichmag}, but none of them are conclusive.

The estimated ejecta mass of OGLE14-073 ($60^{+42}_{-16}~\Msun$) is one of the largest among any SNe currently known. It is presumed that the energy released by the collapse of such massive SN progenitors is too low to unbind the entire stars, and most part of the stellar material falls back to the central remnants, producing black holes \citep[BHs; e.g.,][]{sukhbold2016neutrinoexpl}. This fallback accretion has been suggested to be able to power SNe like superluminous SNe \citep[e.g.,][]{dexter2013fallback}. Given the huge ejecta mass estimated for OGLE14-073, it is possible that initially its massive progenitor failed to explode in a standard way. However, the following fallback accretion powered a luminous transient, which meant that it was observed as an energetic SN. In this Letter, we investigate this scenario as a possible explanation of the peculiar properties of OGLE14-073.

\section{Fallback accretion}
We first investigate the fallback accretion rate that is essential in providing the SN luminosity in our model. We use the semi-analytic approach that is adopted by \citet{dexter2013fallback} and showed to match their numerical results.

When an energy \Ein\ is released at the center of a progenitor, the shock velocity ($v_s$) gained at each mass shell of the progenitor can be approximately described as \citep{matzner1999ackee}
\begin{equation}
v_s \simeq 0.794\left(\frac{\Ein}{m}\right)^{0.5}\left(\frac{m}{\rho_0 r_0^3}\right)^{0.19},
\end{equation}
where $m$ is the enclosed mass of the progenitor, $\rho_0$ is the density at the mass shell, and $r_0$ is the radius of the mass shell. The long lasting fallback accretion that can power SNe for a long time is caused by the region where the shock velocity $v_s$ is close to the escape velocity. The fallback time of such region can be analytically estimated by Eq. (3.7) of \citet{chevalier1989fallback}. Using the fallback time of each mass shell, we can estimate the fallback accretion rate corresponding to a given input energy \Ein.

We estimate the fallback accretion rate for the case of two progenitors from \citet{woosley2002progenitor}. We use their 30~\Msun\ and 40~\Msun\ progenitors, having $10^{-4}~Z_\odot$ at ZAMS. The progenitors experience little mass loss and retain almost all the mass at the core collapse. The hydrogen-rich envelope masses and radii of the progenitors at the moment of collapse are 19~\Msun\ and 49~\Rsun\ (30~\Msun) and 25~\Msun\ and 87~\Rsun\ (40~\Msun).

\begin{figure}
	\includegraphics[width=\columnwidth]{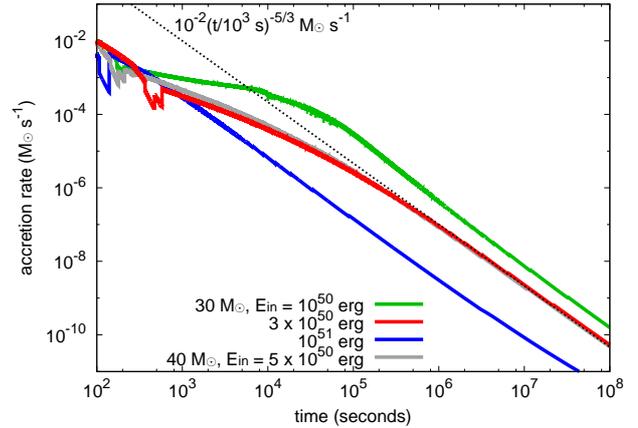}
    \caption{
    Mass accretion rate estimated for the 30~\Msun\ and 40~\Msun\ progenitors. \Ein\ is the initial explosion energy released at the stellar center. The accretion rates eventually follow $\propto t^{-5/3}$ which matches the analytical estimate.
    }
    \label{fig:accrate}
\end{figure}

Fig.~\ref{fig:accrate} shows the estimated accretion rates from the 30~\Msun\ and 40~\Msun\ progenitors. We show the accretion rates corresponding to $\Ein=10^{50}~\mathrm{erg}$, $3\times 10^{50}~\mathrm{erg}$, and $10^{51}~\mathrm{erg}$ for the 30~\Msun\ model. For the 40~\Msun\ progenitor, we show the accretion rate corresponding to $\Ein=5\times 10^{50}~\mathrm{erg}$, which is almost identical to the accretion rate of the 30~\Msun\ progenitor with $\Ein=3\times 10^{50}~\mathrm{erg}$. The accretion rate eventually becomes proportional to $t^{-5/3}$ as expected by the analytical estimates \citep{michel1988fallback,chevalier1989fallback}. The earlier accretion rate is flatter than $\propto t^{-5/3}$, as also found in the study by \citet{dexter2013fallback,zhang2008fallback}. 

The actual energy input \Eacc\ to the ejecta from the accretion is uncertain. The accretion flow is super-Eddington but it is not dense enough to cool by neutrino emission in the long fallback timescale that we are interested in \citep[e.g.,][]{kohri2005accretion}. The accretion flow is radiatively inefficient and optically thick. Such an optically-thick advection-dominated accretion flow can have super-Eddington accretion and can launch a large scale outflow in which \Eacc\ is proportional to the accretion rate $\dot{M}$ \citep[e.g.,][]{dexter2013fallback}
\begin{equation}
\Eacc = \eta \dot{M}c^2,\label{eq:Eacc}
\end{equation}
where $\eta$ is the efficiency factor and $c$ is the speed of light. $\eta$ is estimated to be of the order of $10^{-3}$ \citep{dexter2013fallback} but it is uncertain. Even if the accretion to the central BH launches a jet, the proportionality of \Eacc\ to $\dot{M}$ with $\eta\sim 10^{-3}$ is expected when the energy injection is related to the magnetic field as in the Blandford-Znajek mechanism \citep[e.g.,][]{komissarov2010bzacc}. $\eta\sim 10^{-3}$ is found to match the $\eta$ that is required to explain the late-phase LC of OGLE14-073 (Section~\ref{sec:lightcurves}). If the energy input from the fallback accretion continues, it may push back the matter falling back and the fallback accretion may be weakened or stopped at some moment. The suppression of the fallback accretion leads to the reduction of the heat powering SNe and the SNe could be fainter than predicted in the following section.

\begin{figure}
	\includegraphics[width=\columnwidth]{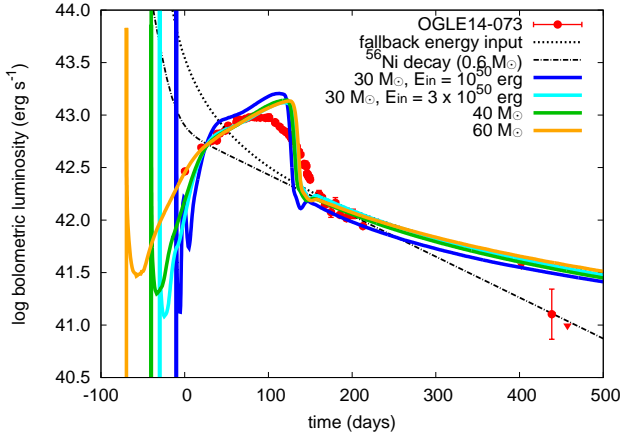}
    \caption{
    Synthetic bolometric LCs. The origin of the time axis is set to the date of the discovery of OGLE14-073. The explosion dates in the synthetic models are $-10$~days (30~\Msun, $\Ein=10^{50}~\mathrm{erg}$), $-30$~days (30~\Msun, $\Ein=3\times 10^{50}~\mathrm{erg}$), $-40$~days (40~\Msun), and $-60$~days (60~\Msun). The dotted line is the energy input from the fallback accretion for the 30~\Msun\ model with $\Ein=3\times10^{50}~\mathrm{erg}$ (Eq.~\ref{eq:Eacc} with $\eta=9\times 10^{-3}$ and the accretion rate corresponding to $\Ein=3\times 10^{50}~\mathrm{erg}$ in Fig.~\ref{fig:accrate}) and its time origin is at $-30$~days. The dot-dashed line shows the nuclear decay energy from 0.6~\Msun\ of \Ni\ which decays as $\Ni\rightarrow\Co\rightarrow\Fe$. Its time origin is $-30$~days. The circles are the bolometric luminosities of OGLE14-073 estimated by \citet{terreran2017ogle73} and the triangles are their upper limits.
    }
    \label{fig:bolLC}
\end{figure}

\begin{figure}
	\includegraphics[width=\columnwidth]{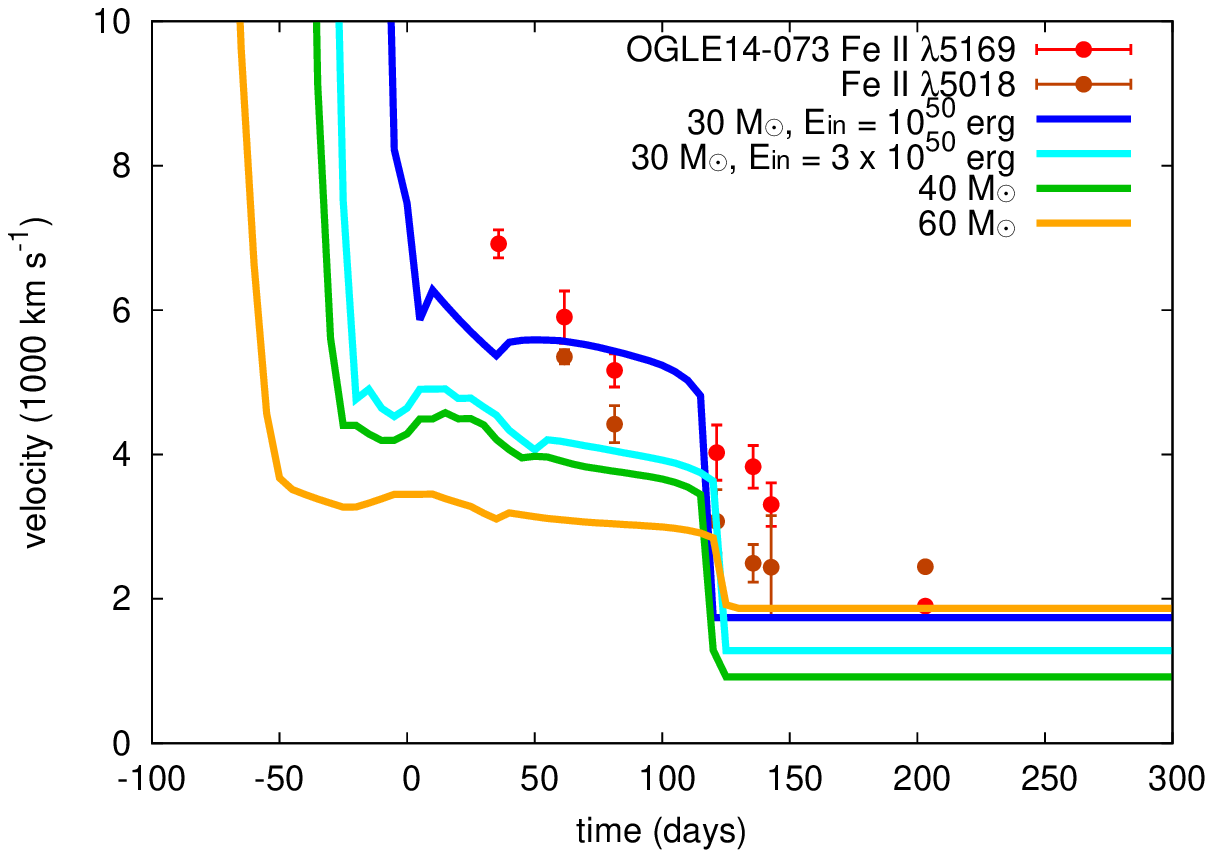}
	\includegraphics[width=\columnwidth]{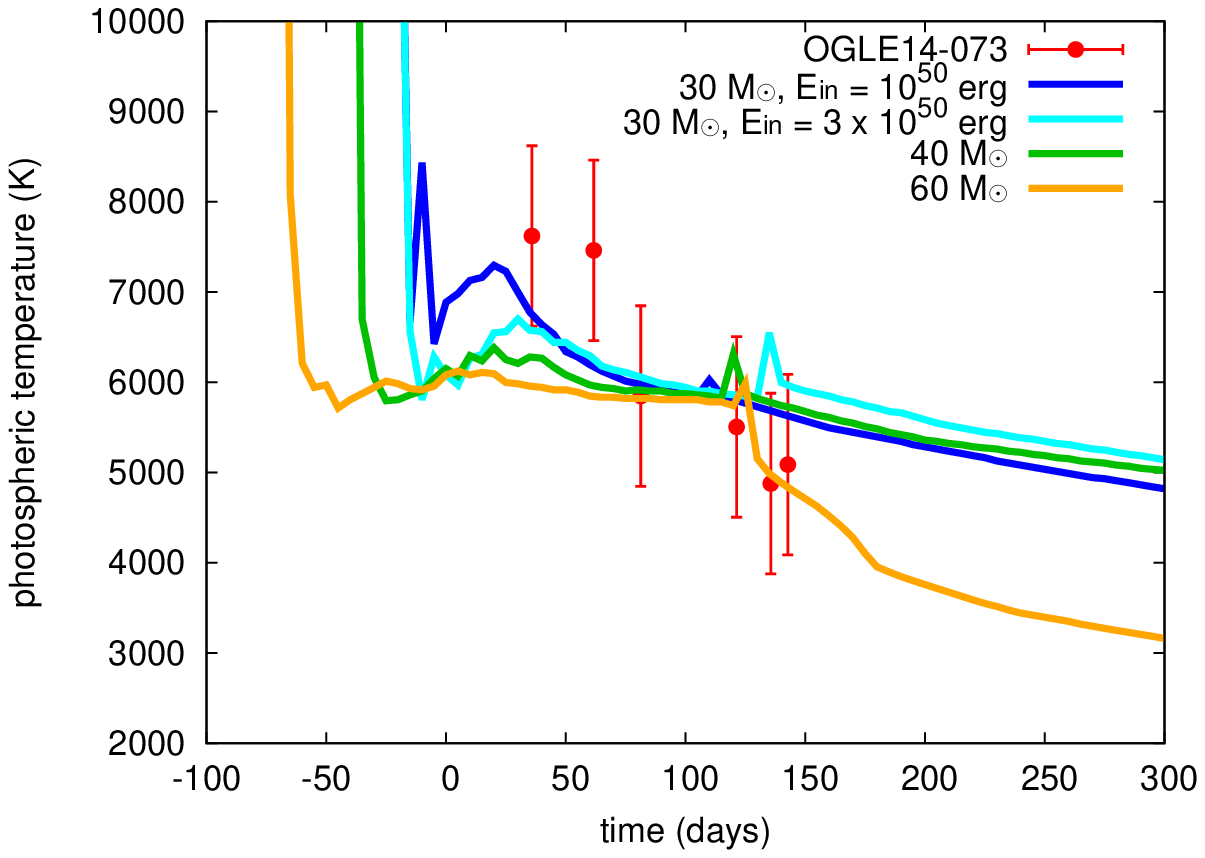}    
    \caption{
    Photospheric velocities (top) and temperatures (bottom) from the fallback accretion powered models and their comparisons to the observations of OGLE14-073 \citep{terreran2017ogle73}. The origin of the time is the same as in Fig.~\ref{fig:bolLC}.
    }
    \label{fig:veltemp}
\end{figure}

\section{Light curves}\label{sec:lightcurves}
We perform numerical LC calculations using a one-dimensional radiation hydrodynamics code \texttt{STELLA} \citep{blinnikov1998sn1993j,blinnikov2000sn1987a,blinnikov2006sniadeflg}. The code evaluates the spectral energy distribution (SED) at each time step and enables us to obtain the time evolution of the luminosities as well as the temperatures and the velocities at the photosphere. 

We first put the 30~\Msun\ progenitor as an initial condition in \texttt{STELLA}. We set the mass cut at 5~\Msun\ to take the fallback into account but the exact location of the mass cut does not affect the LCs significantly when we set it at the order of $1~\Msun$. We adopt the fallback accretion rates corresponding to $\Ein = 10^{50}~\mathrm{erg}$ and $\Ein=3\times 10^{50}~\mathrm{erg}$ shown in Fig.~\ref{fig:accrate}. At the beginning of the calculations, we put \Ein\ as thermal energy just above the mass cut to initiate the initial explosion, which may be caused, for instance, by neutrino heating. Although the central BH mass is expected to increase with time, we set the mass cut at 5~\Msun\ from the beginning to facilitate our numerical calculations. In reality, a proto-neutron star is expected to form at the center during the early phases of the explosion, which then collapses to a BH due to the fallback accretion. Because the LC properties are mainly affected by the final ejecta mass, we start with the large mass cut.

After 100~sec from the initial energy injection, we start to put \Eacc\ (Eq.~\ref{eq:Eacc}) as thermal energy just above the mass cut in which we take $\dot{M}$ from Fig.~\ref{fig:accrate}. At this time, the blast wave from the initial explosion is located in the outer layers, and material has already started to accrete on to the central remnant \citep[e.g.,][]{dexter2013fallback,zhang2008fallback}. Although the fallback accretion may actually begin before 100~sec, the initial energy input does not strongly affect the later LCs we investigate in this paper. Because the thermal energy is always put just below the mass cut, a low density region appears at the central region of the ejecta. We point out that there is no \Ni\ in the models we present, therefore the late-phase LC tail is mainly determined by the central energy input from the fallback accretion, rather than be powered by the decay of \Co\ like in classical SNe. The efficiency $\eta$ is set to match the late-phase LC tail of OGLE14-073. We use $\eta = 2\times 10^{-4}$ ($\Ein=10^{50}~\mathrm{erg}$) and $9\times 10^{-4}$ ($\Ein=3\times 10^{50}~\mathrm{erg}$). We also investigate the 40~\Msun\ and 60~\Msun\ models from \citet{woosley2002progenitor}. In the case of the 40~\Msun progenitor, we adopt the $\Ein=5\times 10^{50}~\mathrm{erg}$ fallback accretion rate of Fig.~\ref{fig:accrate} with $\eta=10^{-3}$. The same accretion rate with $\eta=1.3\times 10^{-3}$ is adopted for simplicity for the 60~\Msun\ model, which has the hydrogen-rich envelope mass of 34~\Msun\ and the radius of 170~\Rsun.

Fig.~\ref{fig:bolLC} shows our synthetic bolometric LCs. The overall LC properties of OGLE14-073 match our fallback accretion powered models. We note that the bolometric LC of OGLE14-073 is constructed based mainly on optical photometric data and the ultraviolet contribution is estimated by extrapolating the optical SEDs assuming a blackbody spectrum.

After the shock breakout, the bolometric LCs quickly decline due to the adiabatic cooling because of the progenitor's relatively small radius. Then, the LCs gradually brighten thanks to injection from the center of the fallback accretion energy. The rise time, which for OGLE14-073 is not well constrained by the observations, depends primarily on the hydrogen-rich envelope mass in the progenitors. The rises of the synthetic LCs last more than 100~days, as observed in OGLE14-073.

Our synthetic LCs are not as \textit{round} as observed in OGLE14-073 and have a spiky peak with an abrupt LC decline. A similar shape is found in the LC models for SN 1987A where no mixing of \Ni\ is considered \citep[e.g.,][]{blinnikov2000sn1987a}. In our models, the heating source is only located at the center. The recombination phase in the hydrogen-rich envelope suddenly ends and the photosphere suddenly recedes to the center. The recession of the photosphere could occur more slowly if we consider the leakage of the accretion energy to more outer layers due to, e.g., collimated ejecta. Collimated ejecta can also induce mixing of \Ni\ outward before falling back. No effect of \Ni\ is considered in our LC models here.

The bolometric luminosities after the LC drop match well the observations for a while. At 400~days, our synthetic LCs are much brighter than observed. The discrepancy could be due to the inefficient fallback accretion at later phases due to the energy injection from the center, for example \citep[e.g.,][]{dexter2013fallback}. The total energy provided to the ejecta through the fallback accretion power (Eq.~\ref{eq:Eacc}) in the 30~\Msun\ model with $\Ein=3\times 10^{50}~\mathrm{erg}$ is $4\times 10^{51}~\mathrm{erg}$. With our assumption of $\eta=9\times 10^{-4}$, it corresponds to 3~\Msun of accreted material.

Fig.~\ref{fig:veltemp} compares photospheric velocities and temperatures of our models to those estimated for OGLE14-073 \citep{terreran2017ogle73}. The photospheric velocity of the 30~\Msun\ with $\Ein=10^{50}~\mathrm{erg}$ matches those of OGLE14-073. The other models predict relatively lower photospheric velocities but the 30~\Msun\ model with $\Ein=3\times 10^{50}~\mathrm{erg}$ and the 40~\Msun\ model are not very far off. The photospheric temperature evolution of the models matches that found in OGLE14-073. Spectroscopic modeling is beyond the scope of this paper but we expect that our models would predict similar spectra to OGLE14-073 given the similar velocities and temperatures found in our LC modeling.

Overall, our synthetic fallback accretion-powered LCs have similar properties to those found in OGLE14-073. The photospheric velocities indicate that the 30~\Msun\ and 40~\Msun\ progenitors better explain OGLE14-037 than the 60~\Msun\ model, which has relatively slow photospheric velocities.



\section{Discussion}
We have shown that the overall observational properties of OGLE14-073 can be reproduced by our fallback accretion-powered model. Failed SNe have been suggested to be observed as faint and less-energetic transients \citep[e.g.,][]{fryer2009failed,moriya2010fallback,lovegrove2013failed}. However, OGLE14-073 may indicate that such failed SNe may actually be observed as energetic SNe. Some of the slowly rising Type~II SNe presented in \citet{taddia2016longrisingtypeii} are also estimated to have similar explosion energies and ejecta masses to OGLE14-073 and this kind of Type~II SNe may not be extremely rare. Meanwhile, the recent discovery of possible disappeared massive stars without explosions \citep{adams2017unnovae} indicate that not all massive stars are powered by the fallback accretion. Some other transients like gamma-ray bursts with extremely long duration may also be related to failed explosions \citep[e.g.,][]{quataert2012kasen}. Further studies are required to see in what conditions the failed explosions can turn into the energetic accretion-powered SNe. 

We made many simplifications in modeling the LCs of the fallback accretion-powered SNe. First of all, the energy produced by the accretion is isotropically released at the center of the progenitor. We also assume that the fallback accretion continues to exist even after a significant amount of energy is released at the center. In a spherically symmetric picture, the fallback accretion energy released at the center can push the accreting matter outwards and stop any further accretion. Again, the fallback accretion may continue for a long time when we take the asphericity into account. As discussed before, the possible sudden luminosity decline at around 400~days may be related to the suppression of the fallback accretion.

The fact that OGLE14-073 matches the fallback accretion-powered model does not exclude other proposed models for OGLE14-073, such as pair-instability SNe. Pair-instability SNe are expected to synthesize a large amount of \Ni, which then decays into \Fe. Therefore, these events could be discerned by some late-time Fe-rich spectra \citep{jerkstrand2016pisnlate}. The magnetar-powered model proposed in \citet{terreran2017ogle73} could be harder to be observationally distinguished from our fallback accretion powered model because they are both caused by the central engine. It is possible that the fallback accretion powered LCs tend to be fainter at late phases because of the suppression of the fallback accretion due to the energy input from inside. Further investigations are required to clarify which observational features are important in distinguishing different suggested models.





\section*{Acknowledgements}
TJM is supported by the Grants-in-Aid for Scientific Research of the Japan Society for the Promotion of Science (16H07413, 17H02864).
Numerical computations were partially carried out on PC cluster at Center for Computational Astrophysics, National Astronomical Observatory of Japan.




\bibliographystyle{mnras}
\bibliography{reference} 

\begin{thebibliography}{}
\makeatletter
\relax
\def\mn@urlcharsother{\let\do\@makeother \do\$\do\&\do\#\do\^\do\_\do\%\do\~}
\def\mn@doi{\begingroup\mn@urlcharsother \@ifnextchar [ {\mn@doi@}
  {\mn@doi@[]}}
\def\mn@doi@[#1]#2{\def\@tempa{#1}\ifx\@tempa\@empty \href
  {http://dx.doi.org/#2} {doi:#2}\else \href {http://dx.doi.org/#2} {#1}\fi
  \endgroup}
\def\mn@eprint#1#2{\mn@eprint@#1:#2::\@nil}
\def\mn@eprint@arXiv#1{\href {http://arxiv.org/abs/#1} {{\tt arXiv:#1}}}
\def\mn@eprint@dblp#1{\href {http://dblp.uni-trier.de/rec/bibtex/#1.xml}
  {dblp:#1}}
\def\mn@eprint@#1:#2:#3:#4\@nil{\def\@tempa {#1}\def\@tempb {#2}\def\@tempc
  {#3}\ifx \@tempc \@empty \let \@tempc \@tempb \let \@tempb \@tempa \fi \ifx
  \@tempb \@empty \def\@tempb {arXiv}\fi \@ifundefined
  {mn@eprint@\@tempb}{\@tempb:\@tempc}{\expandafter \expandafter \csname
  mn@eprint@\@tempb\endcsname \expandafter{\@tempc}}}

\bibitem[\protect\citeauthoryear{{Adams}, {Kochanek}, {Gerke}  \&
  {Stanek}}{{Adams} et~al.}{2017}]{adams2017unnovae}
{Adams} S.~M.,  {Kochanek} C.~S.,  {Gerke} J.~R.,   {Stanek} K.~Z.,  2017,
  \mn@doi [\mnras] {10.1093/mnras/stx898}, \href
  {http://adsabs.harvard.edu/abs/2017MNRAS.469.1445A} {469, 1445}

\bibitem[\protect\citeauthoryear{{Barkat}, {Rakavy}  \& {Sack}}{{Barkat}
  et~al.}{1967}]{barkat1967pisn}
{Barkat} Z.,  {Rakavy} G.,   {Sack} N.,  1967, \mn@doi [Physical Review
  Letters] {10.1103/PhysRevLett.18.379}, \href
  {http://adsabs.harvard.edu/abs/1967PhRvL..18..379B} {18, 379}

\bibitem[\protect\citeauthoryear{{Bersten} \& {Benvenuto}}{{Bersten} \&
  {Benvenuto}}{2016}]{bersten2016hrichmag}
{Bersten} M.~C.,  {Benvenuto} O.~G.,  2016, Boletin de la Asociacion Argentina
  de Astronomia La Plata Argentina, \href
  {http://adsabs.harvard.edu/abs/2016BAAA...58..246B} {58, 246}

\bibitem[\protect\citeauthoryear{{Blinnikov}, {Eastman}, {Bartunov},
  {Popolitov}  \& {Woosley}}{{Blinnikov} et~al.}{1998}]{blinnikov1998sn1993j}
{Blinnikov} S.~I.,  {Eastman} R.,  {Bartunov} O.~S.,  {Popolitov} V.~A.,
  {Woosley} S.~E.,  1998, \mn@doi [\apj] {10.1086/305375}, \href
  {http://adsabs.harvard.edu/abs/1998ApJ...496..454B} {496, 454}

\bibitem[\protect\citeauthoryear{{Blinnikov}, {Lundqvist}, {Bartunov}, {Nomoto}
   \& {Iwamoto}}{{Blinnikov} et~al.}{2000}]{blinnikov2000sn1987a}
{Blinnikov} S.,  {Lundqvist} P.,  {Bartunov} O.,  {Nomoto} K.,   {Iwamoto} K.,
  2000, \mn@doi [\apj] {10.1086/308588}, \href
  {http://adsabs.harvard.edu/abs/2000ApJ...532.1132B} {532, 1132}

\bibitem[\protect\citeauthoryear{{Blinnikov}, {R{\"o}pke}, {Sorokina},
  {Gieseler}, {Reinecke}, {Travaglio}, {Hillebrandt}  \&
  {Stritzinger}}{{Blinnikov} et~al.}{2006}]{blinnikov2006sniadeflg}
{Blinnikov} S.~I.,  {R{\"o}pke} F.~K.,  {Sorokina} E.~I.,  {Gieseler} M.,
  {Reinecke} M.,  {Travaglio} C.,  {Hillebrandt} W.,   {Stritzinger} M.,  2006,
  \mn@doi [\aap] {10.1051/0004-6361:20054594}, \href
  {http://adsabs.harvard.edu/abs/2006A%26A...453..229B} {453, 229}

\bibitem[\protect\citeauthoryear{{Botticella} et~al.,}{{Botticella}
  et~al.}{2010}]{botticella2010sn2009kf}
{Botticella} M.~T.,  et~al., 2010, \mn@doi [\apjl]
  {10.1088/2041-8205/717/1/L52}, \href
  {http://adsabs.harvard.edu/abs/2010ApJ...717L..52B} {717, L52}

\bibitem[\protect\citeauthoryear{{Chevalier}}{{Chevalier}}{1989}]{chevalier1989fallback}
{Chevalier} R.~A.,  1989, \mn@doi [\apj] {10.1086/168066}, \href
  {http://adsabs.harvard.edu/abs/1989ApJ...346..847C} {346, 847}

\bibitem[\protect\citeauthoryear{{Dexter} \& {Kasen}}{{Dexter} \&
  {Kasen}}{2013}]{dexter2013fallback}
{Dexter} J.,  {Kasen} D.,  2013, \mn@doi [\apj] {10.1088/0004-637X/772/1/30},
  \href {http://adsabs.harvard.edu/abs/2013ApJ...772...30D} {772, 30}

\bibitem[\protect\citeauthoryear{{Fryer} et~al.,}{{Fryer}
  et~al.}{2009}]{fryer2009failed}
{Fryer} C.~L.,  et~al., 2009, \mn@doi [\apj] {10.1088/0004-637X/707/1/193},
  \href {http://adsabs.harvard.edu/abs/2009ApJ...707..193F} {707, 193}

\bibitem[\protect\citeauthoryear{{Janka}}{{Janka}}{2012}]{janka2012corecollapsereview}
{Janka} H.-T.,  2012, \mn@doi [Annual Review of Nuclear and Particle Science]
  {10.1146/annurev-nucl-102711-094901}, \href
  {http://adsabs.harvard.edu/abs/2012ARNPS..62..407J} {62, 407}

\bibitem[\protect\citeauthoryear{{Jerkstrand}, {Smartt}  \&
  {Heger}}{{Jerkstrand} et~al.}{2016}]{jerkstrand2016pisnlate}
{Jerkstrand} A.,  {Smartt} S.~J.,   {Heger} A.,  2016, \mn@doi [\mnras]
  {10.1093/mnras/stv2369}, \href
  {http://adsabs.harvard.edu/abs/2016MNRAS.455.3207J} {455, 3207}

\bibitem[\protect\citeauthoryear{{Kohri}, {Narayan}  \& {Piran}}{{Kohri}
  et~al.}{2005}]{kohri2005accretion}
{Kohri} K.,  {Narayan} R.,   {Piran} T.,  2005, \mn@doi [\apj]
  {10.1086/431354}, \href {http://adsabs.harvard.edu/abs/2005ApJ...629..341K}
  {629, 341}

\bibitem[\protect\citeauthoryear{{Komissarov} \& {Barkov}}{{Komissarov} \&
  {Barkov}}{2010}]{komissarov2010bzacc}
{Komissarov} S.~S.,  {Barkov} M.~V.,  2010, \mn@doi [\mnras]
  {10.1111/j.1745-3933.2009.00792.x}, \href
  {http://adsabs.harvard.edu/abs/2010MNRAS.402L..25K} {402, L25}

\bibitem[\protect\citeauthoryear{{Lovegrove} \& {Woosley}}{{Lovegrove} \&
  {Woosley}}{2013}]{lovegrove2013failed}
{Lovegrove} E.,  {Woosley} S.~E.,  2013, \mn@doi [\apj]
  {10.1088/0004-637X/769/2/109}, \href
  {http://adsabs.harvard.edu/abs/2013ApJ...769..109L} {769, 109}

\bibitem[\protect\citeauthoryear{{Matzner} \& {McKee}}{{Matzner} \&
  {McKee}}{1999}]{matzner1999ackee}
{Matzner} C.~D.,  {McKee} C.~F.,  1999, \mn@doi [\apj] {10.1086/306571}, \href
  {http://adsabs.harvard.edu/abs/1999ApJ...510..379M} {510, 379}

\bibitem[\protect\citeauthoryear{{Michel}}{{Michel}}{1988}]{michel1988fallback}
{Michel} F.~C.,  1988, \mn@doi [\nat] {10.1038/333644a0}, \href
  {http://adsabs.harvard.edu/abs/1988Natur.333..644M} {333, 644}

\bibitem[\protect\citeauthoryear{{Moriya}, {Tominaga}, {Tanaka}, {Nomoto},
  {Sauer}, {Mazzali}, {Maeda}  \& {Suzuki}}{{Moriya}
  et~al.}{2010}]{moriya2010fallback}
{Moriya} T.,  {Tominaga} N.,  {Tanaka} M.,  {Nomoto} K.,  {Sauer} D.~N.,
  {Mazzali} P.~A.,  {Maeda} K.,   {Suzuki} T.,  2010, \mn@doi [\apj]
  {10.1088/0004-637X/719/2/1445}, \href
  {http://adsabs.harvard.edu/abs/2010ApJ...719.1445M} {719, 1445}

\bibitem[\protect\citeauthoryear{{Moriya}, {Tominaga}, {Blinnikov}, {Baklanov}
  \& {Sorokina}}{{Moriya} et~al.}{2011}]{moriya2011rsgwindexp}
{Moriya} T.,  {Tominaga} N.,  {Blinnikov} S.~I.,  {Baklanov} P.~V.,
  {Sorokina} E.~I.,  2011, \mn@doi [\mnras] {10.1111/j.1365-2966.2011.18689.x},
  \href {http://adsabs.harvard.edu/abs/2011MNRAS.415..199M} {415, 199}

\bibitem[\protect\citeauthoryear{{Moriya}, {Blinnikov}, {Tominaga}, {Yoshida},
  {Tanaka}, {Maeda}  \& {Nomoto}}{{Moriya} et~al.}{2013}]{moriya2013sn2006gy}
{Moriya} T.~J.,  {Blinnikov} S.~I.,  {Tominaga} N.,  {Yoshida} N.,  {Tanaka}
  M.,  {Maeda} K.,   {Nomoto} K.,  2013, \mn@doi [\mnras]
  {10.1093/mnras/sts075}, \href
  {http://adsabs.harvard.edu/abs/2013MNRAS.428.1020M} {428, 1020}

\bibitem[\protect\citeauthoryear{{Pejcha} \& {Prieto}}{{Pejcha} \&
  {Prieto}}{2015}]{pejcha2015snii}
{Pejcha} O.,  {Prieto} J.~L.,  2015, \mn@doi [\apj]
  {10.1088/0004-637X/806/2/225}, \href
  {http://adsabs.harvard.edu/abs/2015ApJ...806..225P} {806, 225}

\bibitem[\protect\citeauthoryear{{Quataert} \& {Kasen}}{{Quataert} \&
  {Kasen}}{2012}]{quataert2012kasen}
{Quataert} E.,  {Kasen} D.,  2012, \mn@doi [\mnras]
  {10.1111/j.1745-3933.2011.01151.x}, \href
  {http://adsabs.harvard.edu/abs/2012MNRAS.419L...1Q} {419, L1}

\bibitem[\protect\citeauthoryear{{Rakavy} \& {Shaviv}}{{Rakavy} \&
  {Shaviv}}{1967}]{rakavy1967pisn}
{Rakavy} G.,  {Shaviv} G.,  1967, \mn@doi [\apj] {10.1086/149204}, \href
  {http://adsabs.harvard.edu/abs/1967ApJ...148..803R} {148, 803}

\bibitem[\protect\citeauthoryear{{Sukhbold} \& {Thompson}}{{Sukhbold} \&
  {Thompson}}{2017}]{sukbold2017hrichmag}
{Sukhbold} T.,  {Thompson} T.~A.,  2017, \mn@doi [\mnras]
  {10.1093/mnras/stx2004}, \href
  {http://adsabs.harvard.edu/abs/2017MNRAS.472..224S} {472, 224}

\bibitem[\protect\citeauthoryear{{Sukhbold}, {Ertl}, {Woosley}, {Brown}  \&
  {Janka}}{{Sukhbold} et~al.}{2016}]{sukhbold2016neutrinoexpl}
{Sukhbold} T.,  {Ertl} T.,  {Woosley} S.~E.,  {Brown} J.~M.,   {Janka} H.-T.,
  2016, \mn@doi [\apj] {10.3847/0004-637X/821/1/38}, \href
  {http://adsabs.harvard.edu/abs/2016ApJ...821...38S} {821, 38}

\bibitem[\protect\citeauthoryear{{Taddia} et~al.,}{{Taddia}
  et~al.}{2016}]{taddia2016longrisingtypeii}
{Taddia} F.,  et~al., 2016, \mn@doi [\aap] {10.1051/0004-6361/201527811}, \href
  {http://adsabs.harvard.edu/abs/2016A%26A...588A...5T} {588, A5}

\bibitem[\protect\citeauthoryear{{Terreran} et~al.,}{{Terreran}
  et~al.}{2017}]{terreran2017ogle73}
{Terreran} G.,  et~al., 2017, \mn@doi [Nature Astronomy]
  {10.1038/s41550-017-0228-8}, \href
  {http://adsabs.harvard.edu/abs/2017NatAs...1E.228T} {1, 228}

\bibitem[\protect\citeauthoryear{{Utrobin}, {Chugai}  \&
  {Botticella}}{{Utrobin} et~al.}{2010}]{utrobin2010sn2009kf}
{Utrobin} V.~P.,  {Chugai} N.~N.,   {Botticella} M.~T.,  2010, \mn@doi [\apjl]
  {10.1088/2041-8205/723/1/L89}, \href
  {http://adsabs.harvard.edu/abs/2010ApJ...723L..89U} {723, L89}

\bibitem[\protect\citeauthoryear{{Woosley}}{{Woosley}}{2017}]{woosley2017ppisn}
{Woosley} S.~E.,  2017, \mn@doi [\apj] {10.3847/1538-4357/836/2/244}, \href
  {http://adsabs.harvard.edu/abs/2017ApJ...836..244W} {836, 244}

\bibitem[\protect\citeauthoryear{{Woosley}, {Heger}  \& {Weaver}}{{Woosley}
  et~al.}{2002}]{woosley2002progenitor}
{Woosley} S.~E.,  {Heger} A.,   {Weaver} T.~A.,  2002, \mn@doi [Reviews of
  Modern Physics] {10.1103/RevModPhys.74.1015}, \href
  {http://adsabs.harvard.edu/abs/2002RvMP...74.1015W} {74, 1015}

\bibitem[\protect\citeauthoryear{{Zhang}, {Woosley}  \& {Heger}}{{Zhang}
  et~al.}{2008}]{zhang2008fallback}
{Zhang} W.,  {Woosley} S.~E.,   {Heger} A.,  2008, \mn@doi [\apj]
  {10.1086/526404}, \href {http://adsabs.harvard.edu/abs/2008ApJ...679..639Z}
  {679, 639}

\makeatother
\end{thebibliography}






\bsp	
\label{lastpage}
\end{document}